\documentclass[12pt]{article}
\usepackage{graphicx}

\usepackage{amsmath}
\usepackage{amssymb}

\usepackage{fullpage}
\usepackage{color}

\def \sixbox {\rule{6pt}{6pt}}

\begin{document}

\title{Adaptive behavior can produce maladaptive anxiety due to individual differences in experience}
\author{Frazer Meacham$^1$ \and Carl Bergstrom$^1$}

\maketitle 

\begin{center}
$^1$Department of Biology, University of Washington\\
Box 351800, Seattle, WA 98195\\
\end{center}

\begin{abstract}

Normal anxiety is considered an adaptive response to the possible presence of 
danger, but is susceptible to dysregulation. Anxiety disorders are 
prevalent at high frequency in contemporary human societies, yet impose 
substantial disability upon their sufferers. This raises a puzzle: why has evolution 
left us vulnerable to anxiety disorders? We develop a signal detection model in
which individuals must learn how to calibrate their anxiety responses: they need
to learn which cues indicate danger in the environment. We derive the optimal strategy for doing so, and find that individuals face an inevitable exploration-exploitation tradeoff between obtaining a better estimate of the level of risk on one hand, and maximizing current payoffs on the other. Because of this tradeoff, a subset of the population can become trapped in a state of self-perpetuating over-sensitivity to threatening stimuli, even when individuals learn optimally. This phenomenon arises because when individuals become too cautious, they stop sampling the environment and fail to correct their misperceptions, whereas when individuals become too careless they continue to sample the environment and soon discover their mistakes. Thus, over-sensitivity to threats becomes common whereas under-sensitivity becomes rare. We suggest that this process may be involved in the development of excessive anxiety in humans.
\end{abstract}

\begin{center}
\end{center}

\section{Introduction}

Motile animals have evolved elaborate mechanisms for detecting and avoiding danger. Many of these mechanisms are deeply conserved evolutionarily \cite{mendl2010integrative}. 
When an individual senses possible danger, this triggers a cascade of physiological responses that prepare it to deal with the threat. 
Behavioral ecological models treat the capacity for anxiety as a mechanism of regulating how easily these defensive responses are induced \cite{marks_fear_1994,nesse2001smoke,nesse_natural_2005,hinds_psychology_2010, bateson2011anxiety,nettle2012evolutionary}. Greater anxiety causes an individual to be alert to more subtle signs of potential danger, while lowered anxiety causes the individual to react only to more obvious signs \cite{burman_anxiety-induced_2009}. As unpleasant as the experience of anxiety may be, the capacity for anxiety is helpful in tuning behavior to environmental circumstance. This viewpoint is bolstered by epidemiological evidence suggesting that long-term survival is worse for people with low anxiety-proneness than for those in the middle of the distribution, due in part to increased rates of accidents and accidental death in early adulthood \cite{lee_protective_2006, mykletun_levels_2009}.

While the capacity for anxiety is adaptive, dysregulated anxiety is also common, at least in humans. Of all classes of mental disorders, anxiety disorders affect the largest number of patients \cite{kessler2009global}. The global prevalence of individuals who suffer from an anxiety disorder at some point in their life is commonly estimated at around 15 percent \cite{kessler2009global, somers2006prevalence}, with 5 to 10 percent of the population experiencing pathological anxiety in any given year \cite{kessler2009global, somers2006prevalence, baxter_global_2013}. The consequences can be drastic: in a 12 month period in the US, 4 percent of individuals had an anxiety disorder that was severe enough to cause work disability, substantial limitation, or more than 30 days of inability to maintain their role \cite{kessler_rc_prevalence_2005}.
The prevalence and magnitude of anxiety disorders is also reflected in the aggregate losses they cause to economic productivity: in the 1990s the annual cost was estimated at \$42 billion in the US alone \cite{greenberg_economic_1999}.

Episodes of clinically-significant anxiety are distributed broadly across the lifespan, and anxiety disorders typically manifest before or during the child-rearing years \cite{kessler2005lifetime}.
Because of the severity of impairment that often results from anxiety disorders, and the fact that onset occurs before or during reproduction, these disorders will often have a substantial effect on Darwinian fitness.
Thus, the prevalence of anxiety disorders poses an apparent problem for the evolutionary viewpoint. If the capacity for anxiety is an adaptation shaped by natural selection, why is it so prone to malfunction?

One possible explanation invokes the so-called smoke detector principle \cite{nesse2001smoke,nesse_natural_2005}. The basic idea is to think about how anxiety serves to help an organism detect danger, and to note the asymmetry between the low cost of a false alarm and the high cost of failing to detect a true threat. This allows us to frame anxiety in the context of signal detection theory. Because of asymmetry in costs of false alarms versus false complacency, the theory predicts that optimized warning systems will commonly generate far more false positives than false negatives. This provides an explanation for why even optimal behavior can produce seemingly excessive sensitivity in the form of frequent false alarms \cite{nesse_why_1994, nesse_natural_2005}. More recently, the signal detection framework has been expanded to describe how the sensitivity of a warning system should track a changing environment and become more easily triggered in dangerous situations \cite{nettle2012evolutionary}. This approach,
together with error management theory \cite{johnson2013evolution},
begins to provide an account of how anxiety and mood regulate behavior over time, and why high levels of anxiety may be adaptive even when true threats are scarce. Better to be skittish and alive than calm but dead.  

The smoke detector principle cannot be the whole story, however. There are a number of aspects of anxiety that it does not readily explain. First, the smoke detector principle deals with evolutionarily adaptive anxiety --- but not with the issue of why evolution has left us vulnerable to anxiety {\em disorders}.  A fully satisfactory model of anxiety and anxiety disorders should explain within-population variation: Why does a small subset of the population suffer from an
excess of anxiety, while the majority regulate anxiety levels appropriately? Second, a critical component of anxiety disorders is the way they emerge from self-reinforcing negative behavior patterns. Individuals with anxiety disorders often avoid situations or activities that are in fact harmless or even beneficial. Effectively, these individuals are behaving too pessimistically, treating harmless situations as if they were dangerous. We would like to explain how adaptive behavior might lead to self-reinforcing pessimism.
Third, if the evolutionary function of anxiety is to modulate the threat response according to environmental circumstances \cite{nesse2009evolution},  evolutionary models of anxiety will need to explicitly treat that modulation process---that is, such models should incorporate the role of learning explicitly.

In this paper, we show that optimal learning can generate behavioral over-sensitivity to threat that is truly harmful to the individual's fitness,
but expressed in only a subset of the population.  Our aim is not to account for the specific details of particular anxiety disorders---phobias, generalized anxiety disorder, post-traumatic stress disorder, and so forth---but rather to capture some of the general features of how anxiety is regulated and how this process can go awry.

In section 2, we illustrate the basic mechanism behind our result using a very simple model borrowed from foraging theory \cite{mcnamara_application_1980} in which an actor must learn by iterative trial and error whether taking some action is unacceptably dangerous or sufficiently safe. (Trimmer et al. \cite{trimmer2015depression} independently developed a related model to study clinical depression. Also see Frankenhuis and Panchanathan \cite{frankenhuis2011balancing} as well as \cite{panchanathan2016evolution} for closely related models of developmental plasticity in general.)
In section 3, we extend the model into the domain of signal detection theory and consider how an actor learns to set the right threshold for responding to an indication of danger.  In most signal detection models, the agent making the decision is assumed to know the distribution of cues generated by safe and by dangerous situations. But where does this knowledge come from? Unless the environment is homogeneous in time and space over evolutionary timescales, the distributions of cues must be learned. In our model, therefore, the agent must actively learn how the cues it observes relate to the presence of danger. We show that under these circumstances, some members of a population of optimal learners will become overly pessimistic in their interpretations of cues, but fewer will become overly optimistic.

\section{Learning about an uncertain world}

If we want to explain excess anxiety
from an evolutionary perspective, we must account for why only a subset of the population is affected. Although genetic differences may be partly responsible, random variation in individual experience can also lead to behavioral differences among individuals.
In particular, if an individual has been unfortunate during its early experience, it may become trapped in a cycle of self-reinforcing pessimism.
To demonstrate this, we begin with a simple model that shows how responses to uncertain conditions are shaped by individual learning.
The model of this section does not include the possibility of the individual observing cues of the potential danger. Thus, it does not capture anxiety's essential characteristic of threat detection. But this model does serve to illustrate the underlying mechanism that can lead a subset of the population to be overly pessimistic.

\subsection{Model}

Because our aim is to reveal general principles around learned pessimism, rather than to model specific human pathologies, we frame our model as a simple fable. Our protagonist is a fox. In the course of its foraging, it occasionally comes across a burrow in the ground. Sometimes the burrow will contain a rabbit that the fox can catch and eat, but sometimes the burrow will contain a fierce badger that may injure the fox. Perhaps our fox lives in an environment where badgers are common, or perhaps it lives in an environment where badgers are rare, but the fox has no way of knowing beforehand which is the case. Where badgers are rare, it is worth taking the minor risk involved in digging up a burrow to hunt rabbits. Where badgers are common, it is not worth the risk and the fox should eschew burrows in favor of safer foraging options: mice, birds, fruits, berries etc. The fox encounters burrows one at a time, and faces the decision of whether to dig at the burrow or whether to slink away. The only information available to the fox at each decision point is the prior probability that badgers are common, and its own experiences with previous burrows.

To formalize this decision problem, we imagine that the fox encounters a sequence of burrows, one after the other. The fox makes a single decision of whether to explore each burrow before encountering the next burrow, and each burrow contains either a rabbit or a badger.
We let $R$ be the payoff to the fox for digging up a burrow that contains a rabbit and $C$ be the cost of digging up a burrow that contains a badger. If the fox decides to leave a burrow undisturbed, its payoff is zero. When the fox decides to dig up a burrow, the probability of finding a badger is $p_g$ if badgers are rare, and $p_b$ if badgers are common, where $p_g < p_b$. If badgers are rare it is worthwhile for the fox to dig up burrows, in the sense that the expected payoff for digging is greater than zero. That is, we assume that
\[
(1-p_g)R - p_gC > 0.
\]
If badgers are common, burrows are best avoided, because the expected payoff for digging is less than zero:
\[
(1-p_b)R - p_bC < 0.
\]
We let $q_0$ be the prior probability that badgers are common and we assume that the correct prior probability is known to the fox. We assume a constant extrinsic death rate $d$ for the fox (and we assume that badger encounters are costly but not lethal), so that the present value of future rewards is discounted by $\lambda = 1-d$ per time step.

If the fox always encountered only a single burrow in its lifetime, calculating the optimal behavior would be straightforward. If the expected value of digging exceeds the expected value of not doing so, the fox should dig. That is, the fox should dig when
\[
(1-q_0)\bigl((1-p_g)R - p_gC)\bigr)+q_0\bigl((1-p_b)R - p_bC\bigr) > 0.
\]
But the fox will very likely encounter a series of burrows, and so as we evaluate the fox's decision at each stage we must also consider the value of the information that the fox gets from digging. Each time the fox digs up a burrow, it gets new information: did the burrow contain a rabbit or a badger?  Based on this information, the fox can update its estimate of the probability that the environment is favorable. If the fox chooses not to dig, it learns nothing and its beliefs remain unchanged. Thus even if the immediate expected value of digging at the first burrow is less than 0, the fox may still benefit from digging because it may learn that the environment is good and thereby benefit substantially from digging at subsequent burrows. In other words, the fox faces an exploration-exploitation tradeoff \cite{kaelbling1996reinforcement} in its decision about whether to dig or not. Because of this tradeoff, the model has the form of a one-armed bandit problem \cite{bertsekas_dynamic_2012}, where the bandit arm returns a payoff of either $R$ or $-C$, and the other arm always returns a payoff of zero. 

\subsection{Optimal behavior}

As an example, suppose good and bad environments are equally likely \emph{a priori} ($q_0=0.5$) and foxes die at a rate of $d=0.05$ per time step.
For simplicity we set the costs and rewards to be symmetric: $C=1$, $R=1$, $p_g=1/4$, $p_b=3/4$. In a good environment where badgers are less common, the expected value of digging up a burrow is positive ($-0.25+(1-0.25)=0.5$) whereas in a bad environment where badgers are common, the expected value of digging up a burrow is negative ($-0.75+(1-0.75)=-0.5$). (Recall that the fox also has other foraging options available, and therefore will not necessarily starve if it avoids the burrows.)

\begin{figure}
  \includegraphics[width=\linewidth]{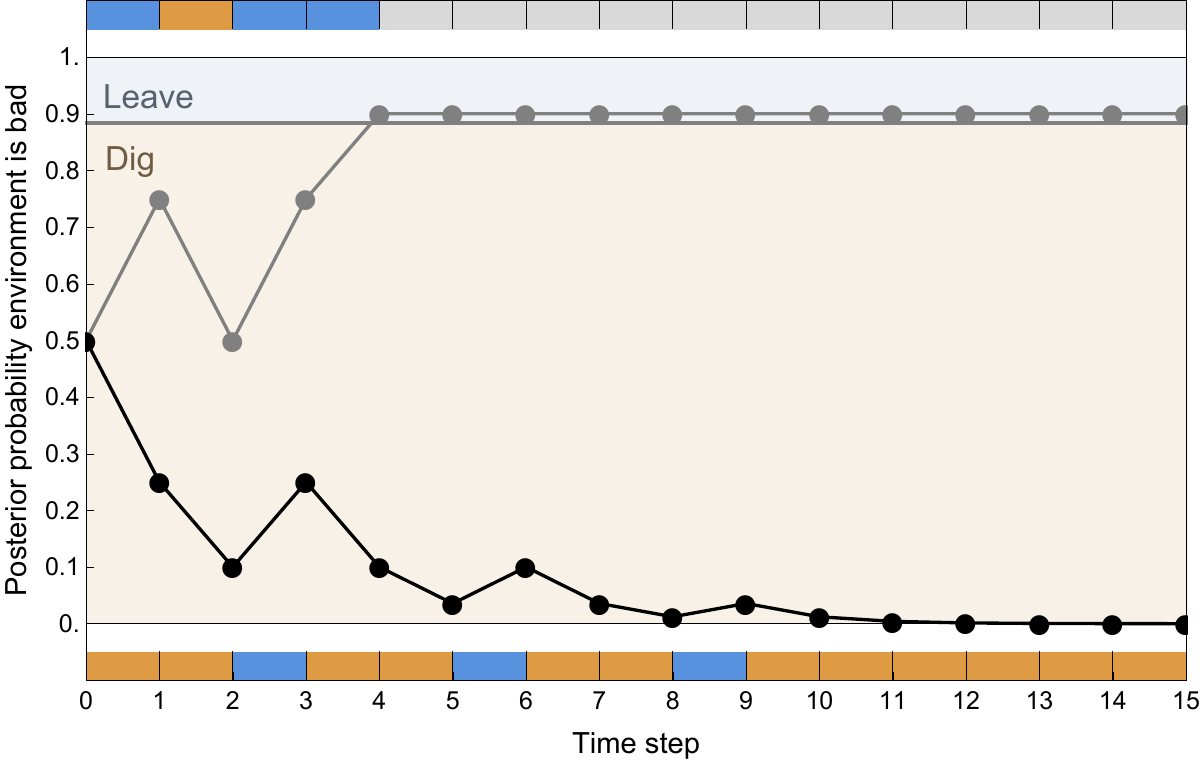}
  \caption{Two examples of optimal behavior by the fox. The vertical axis indicates the fox's posterior subjective probability that it is in a bad environment. In the tan region, the fox should dig. In the blue region, the fox should avoid the burrow. The grey path and black path trace two possible outcomes of a fox's foraging experience. The colored bars above and below the graph indicate the fox's experience along the upper and lower paths respectively: brown indicates that the fox found a rabbit and blue indicates that the fox found a badger. Along the grey path, the fox has a few bad experiences early. This shifts the fox's subjective probability that the environment is bad upward, into the blue region. The fox stops sampling, its probability estimate stays fixed, and learning halts. Along the black path, the fox finds two or more rabbits between each encounter with a badger. Its subjective probability remains in the tan zone throughout, and the fox continues to sample---and learn---throughout the experiment.}
  \label{fig:traces}
\end{figure}

Applying dynamic programming to this scenario (see Appendix B), we find that the fox's optimal behavior is characterized by a threshold value of belief that the environment is bad, above which the fox does not dig at the burrows. (This threshold is the same at all time steps.) Figure \ref{fig:traces} illustrates two different outcomes that a fox might experience when using this optimal strategy. Along the upper path, shown in gray, a fox initially encounters a badger. This is almost enough to cause the fox to conclude he is in a bad environment and stop sampling. But not quite---the fox samples again, and this time finds a rabbit. In his third and fourth attempts, however, the fox encounters a pair of badgers, and that's enough for him---at this point he does give up. Since he does not sample again, he gains no further information and his probability estimate remains unchanged going forward. Along the lower path, shown in black, the fox initially encounters a series of rabbits, and his probability estimate that he is in a bad environment becomes quite low. Even the occasional encounter with a badger does not alter this probability estimate enough that the fox ought to stop sampling, so he continues to dig at every hole he encounters and each time adjusts his probability estimate accordingly.

\subsection{Population outcomes}

After solving for the optimal decision rule, we can examine statistically what happens to an entire population of optimally-foraging foxes. To see what the foxes have learned, we can calculate the population-wide distribution of individual subjective posterior probabilities that the environment is bad. We find that almost all of the foxes who are in unfavorable environments correctly infer that things are bad, but a substantial minority of foxes in favorable circumstances fail to realize that things are good. In Appendix A we show that the general pattern illustrated here is generally robust to variation in model parameters.

\begin{figure}
\begin{center}
\includegraphics[width=10cm]{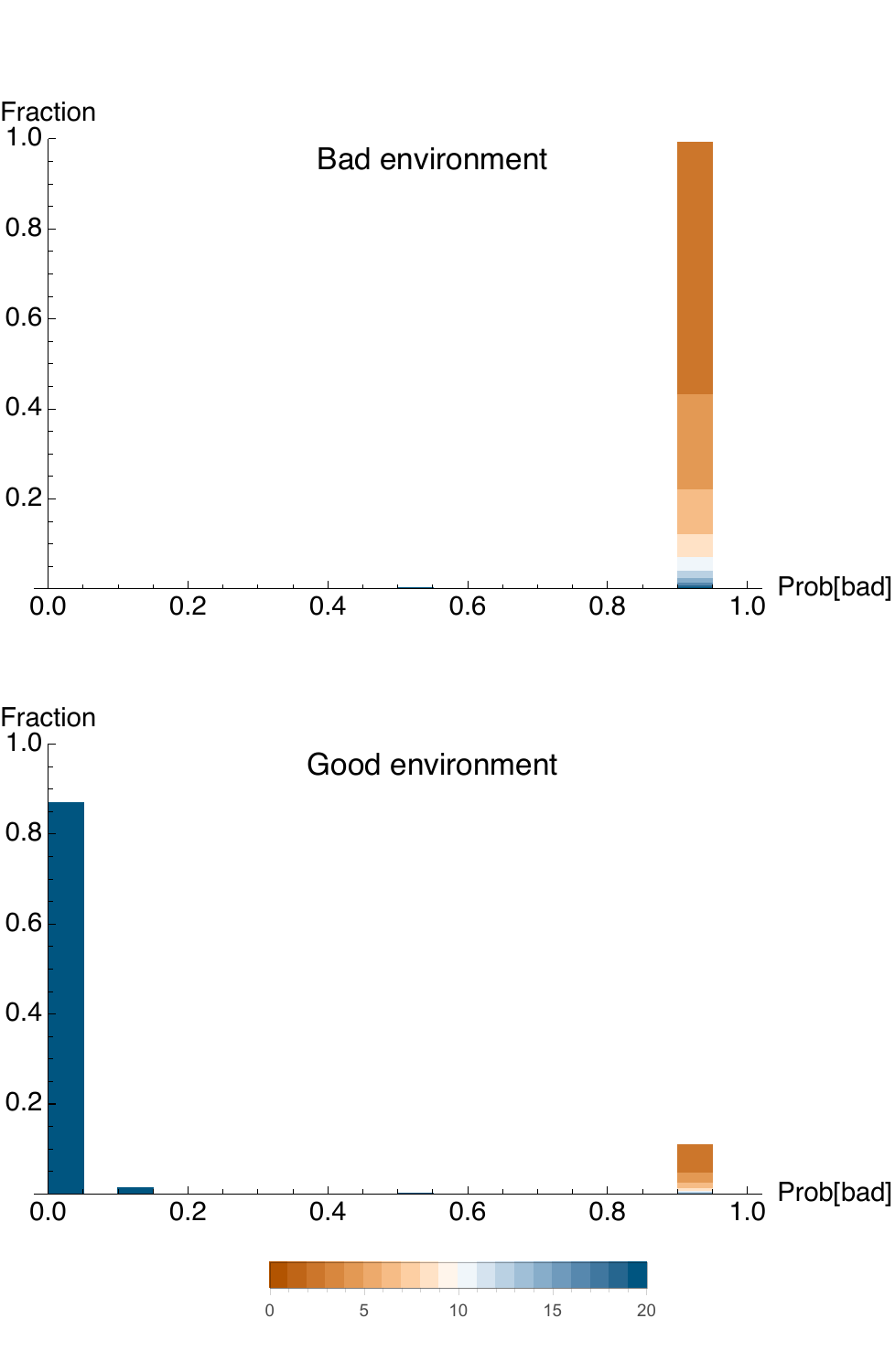}
\end{center}
  \caption{Population distribution of individual posterior probabilities that the environment is bad when the environment is indeed bad (upper panel), and when the environment is actually good (lower panel). The horizontal axis is the individual's posterior probability estimate that environment is bad after 20 opportunities to dig at a burrow. (This is among foxes who have lived that long. Conditioning in this way introduces no sampling bias because survival is independent of environment and behavior in the model.) Frequency is plotted on the vertical axis. Color indicates the number of times an individual has sampled the environment. All individuals began with a prior probability of 0.5 that the environment is bad. When the environment is indeed bad, only 0.2\% of the population erroneously believe the environment is likely to be good. When the environment is good, 11.1\% of the population erroneously believes that it is likely to be bad. The majority of these individuals have sampled only a few times and then given up after a bit of bad luck.
}
  \label{fig:simplemodel}
\end{figure}

Figure \ref{fig:simplemodel} shows the distribution of posterior subjective probabilities that the environment is good among a population of optimally learning foxes for the above parameter choices. We can see that a non-negligible number of individuals in the favorable environment come to the false belief that the environment is probably bad. This occurs because even in a favorable environment, some individuals will uncover enough badgers early on that it seems to them probable that the environment is unfavorable. When this happens those individuals will stop digging up burrows. They will therefore fail to gain any more information, and so their pessimism is self-perpetuating. 

\subsection{Comments}

This self-perpetuating pessimism is not a consequence of a poor heuristic for learning about the environment; we have shown that this phenomenon occurs when individuals are using the {\em optimal} learning strategy. Because of the asymmetry of information gain between being cautious and being exploratory, there results an asymmetry in the numbers of individuals who are overly pessimistic versus overly optimistic. Even when individuals follow the optimal learning rule, a substantial subset of the population becomes too pessimistic but very few individuals become too optimistic.

One might think, knowing that the current learning rule leads to excessive pessimism on average, that we could do better on average by altering the learning rule to be a bit more optimistic. This is not the case. Any learning rule that is more optimistic will result in lower expected payoffs to the learners, and thus would be replaced under natural selection by our optimal learning rule. 

This scenario may reflect an important component of pathological human pessimism or anxiety. For example, many people think that they ``can't sing" or ``are no good at math" because early failures, perhaps during childhood, led to beliefs that have never been challenged. When someone believes he can't sing, he may avoid singing and will therefore never have the chance to learn that his voice is perfectly good. Thus, attitudes that stem from earlier negative experiences become self-perpetuating.

\section{Modeling anxiety by including cues}

In the model we have just explored, the fox knows nothing about a new burrow beyond the posterior probability it has inferred from its past experience. In many situations, however, an individual will be able to use additional cues to determine the appropriate course of action. For example, a cue of possible danger, such as a sudden noise or looming object, can trigger a panic or flight response, and anxiety can be seen as conferring a heightened sensitivity to such signs of threat.
In this view, the anxiety level of an individual determines its sensitivity to indications of potential danger. The higher the level of anxiety, the smaller the cue needed to trigger a flight response  \cite{nesse2001smoke, nesse_natural_2005, bateson2011anxiety, nettle2012evolutionary}.
To model anxiety in this sense, we extend our model of fox and burrow to explore how individuals respond to signs of potential threat.  We will find that even with the presence of cues, a substantial fraction of individuals will fall into a self-perpetuating pattern where their anxiety levels are set too high.

The key consideration in our model is that individuals must {\em learn} how cues correspond to potential threats. In other words, individuals need to calibrate their responses to environmental cues, setting anxiety levels optimally to avoid predators without wasting too much effort on unnecessary flight. Admittedly, if the environment is homogeneous in space and extremely stable over many generations, then natural selection may be able to encode the correspondence between cues and danger into the genome. But when the environment is less predictable, the individual faces the problem of learning to properly tune its responses to cues of possible threat.

\subsection{Model}

We return to our story of the fox, who we now suppose can listen at the entrance to the burrow before deciding whether to dig it up. Rabbits typically make less noise than badgers, so listening can give the fox a clue as to the contents of the burrow. When the burrow is relatively silent it is more likely to contain a rabbit, and when the fox hears distinct snuffling and shuffling noises it is likely that the burrow contains a badger. But the sounds aren't fully reliable. Sometimes rabbits can be noisy, and sometimes badgers are quiet. So although the amount of noise coming from the burrow gives the fox some information about how likely the burrow is to contain a badger, the information is probabilistic and the fox can never be certain.

In contrast to the model of the previous section, the difference between environments is now a matter of how easy it is for the fox to distinguish between dangerous and safe situations, rather than how common danger is. If the environment is good, the fox only needs to be cautious if a burrow is quite noisy. But if the environment is bad, then the fox should be cautious even if faint noises emanate from a burrow. This is because when the environment is bad, it is too risky to dig up a burrow unless the burrow is nearly silent. The fox does not know beforehand whether the environment is good or bad, and therefore it does not know how the probability of finding a badger in the burrow depends on the amount of noise it hears. The only way for it to gain information is to learn by experience.

To formalize the problem, we extend the model in section 2 by supposing that the fox observes a cue before each decision. The cue is a continuous random variable drawn from Gaussian distributions that depend on the environment and what is in the burrow.
We first consider the good environment. As before, we let $p_g$ be the probability that any given burrow contains a badger. When the burrow contains a badger, the cue strength is drawn from a Gaussian distribution with mean $\mu_{g,c}$ and standard deviation $\sigma_{g,c}$. When the burrow contains a rabbit, the cue strength is drawn from a Gaussian distribution with mean $\mu_{g,r}$ and standard deviation $\sigma_{g,r}$.
Similarly for the bad environment, we let $p_b$ be the probability that any given burrow contains a badger, with a cue strength drawn from a Gaussian distribution with mean $\mu_{b,c}$ and standard deviation $\sigma_{b,c}$ when the burrow contains a badger, and mean $\mu_{b,r}$ and standard deviation $\sigma_{b,r}$ when the burrow contains a rabbit.

After observing the cue, the fox decides whether to dig or leave. If the fox decides to leave, its payoff is zero. As before, the cost of encountering a badger is $C$ and the reward for finding a rabbit is $R$. The prior probability that the environment is bad is $q_0$ and future decisions are discounted at a rate of $\lambda$ per time step.
Although not as simple as before, we can again use dynamic programming to calculate the optimal behavior (see Appendix B).

\subsection{Optimal behavior}

In this extended model, the good and bad environments can differ not only in the frequency of badgers, but also in how readily badgers can be distinguished from rabbits by sound alone. Here we will investigate what happens when in good environments, badgers are much louder than rabbits, but in bad environments they are only a little bit louder.
We are particularly interested in this case because we want to know what happens when the fox must learn how cues correspond to potential threats.

\begin{figure}
\includegraphics[width=15cm]{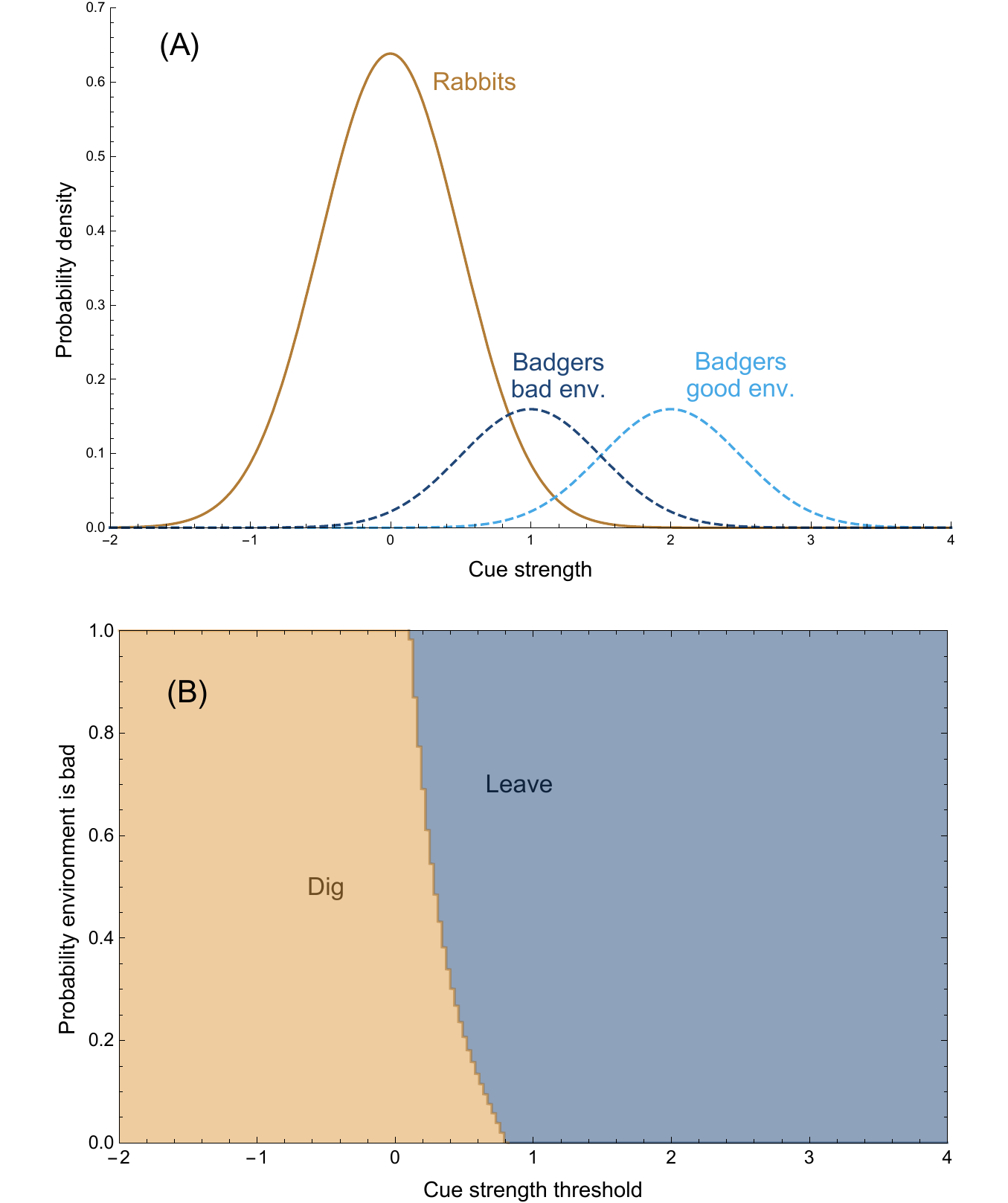}
\caption{
The two environments differ in how loud badgers are (A). In the good environment, badgers are easier to detect than they are in the bad environment. The optimal decision rule is computed using dynamic programming and illustrated in the lower panel (B). The decision about whether to dig depends on the value $x$ of the cue and the subjective probability that the environment is bad. A curve separates the region in which one should dig (tan) from the region in which one should not (blue).}
  \label{fig:complexstrategy}
\end{figure}

To model this situation, we set the mean loudness of rabbits to 0 in both good and bad environments ($\mu_{g,r} = \mu_{b,r} = 0$). (The scale is arbitrary; we have chosen the value 0 for convenience.) In the good environment, badgers are much louder than rabbits ($\mu_{g,c} = 2$), and are therefore usually easy to detect. In the bad environment, they are only a bit louder than rabbits ($\mu_{b,c} = 1$) which can make them more difficult to detect. Everything else about the signal detection problem in the two environments is the same: $\sigma_{g,r} = \sigma_{g,c} = \sigma_{b,r} = \sigma_{b,c} = 0.5$, and $p_g = p_b = 0.2$. Figure \ref{fig:complexstrategy}A shows the distributions of cue intensities for the two environments. The punishment for encountering a badger is greater than the reward for finding a rabbit ($R = 1$, $C = 19$) and as in the previous model, future rewards are discounted at a rate of $\lambda = 0.95$ per time step and good and bad environments are equally common ($q_0 = 0.5$).

The optimal decision rule for the fox, as found by dynamic programming, is illustrated in Figure \ref{fig:complexstrategy}B. The fox now takes into account both its subjective probability that the environment is bad and the intensity of the cue it observes. A curve separates the (cue, probability) pairs at which the fox should dig from the (cue, probability) pairs at which the fox should not. For cues below 0.11, the fox should dig irrespective of the state of the environment; for cues above 0.81, the fox should not dig under any circumstance. In between, the fox must balance the strength of the cue against its subjective probability that the environment is bad.
Here we can see the exploration-exploitation tradeoff in action.
Given the large payoff to be gained from exploiting a good environment over many time steps, the possibility of discovering that the environment is good may compensate for the risk of punishment---even when it is more likely than not that the environment is bad.

\begin{figure}
\includegraphics[width=6in]{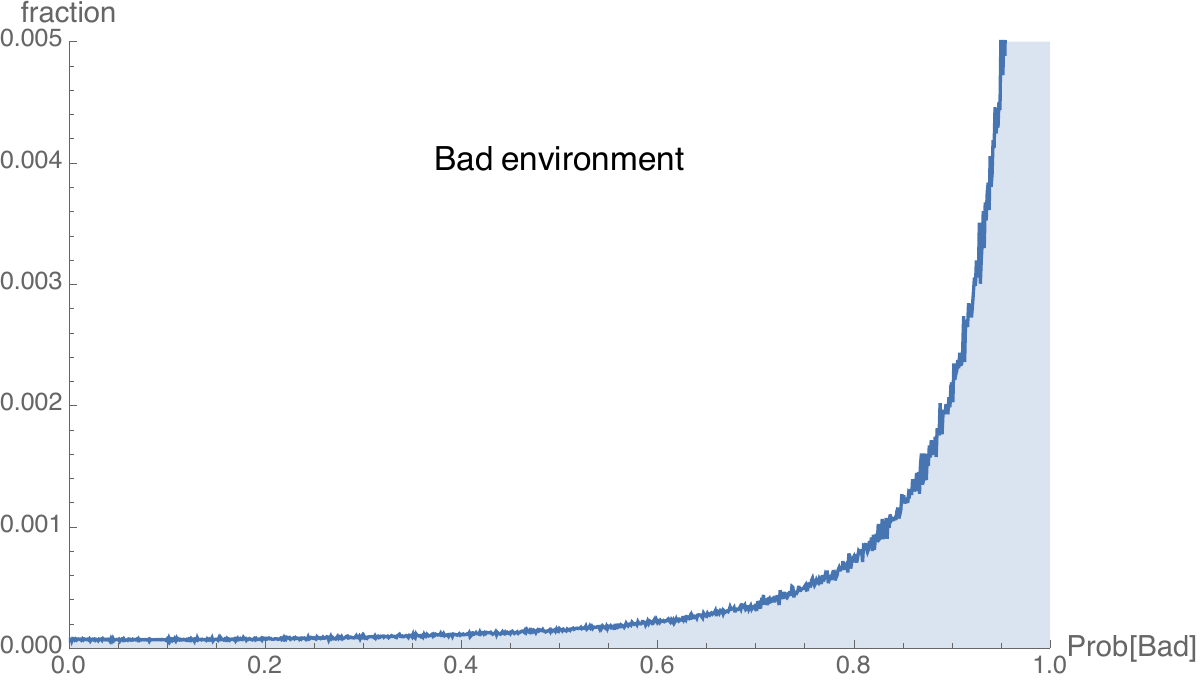}
\includegraphics[width=6in]{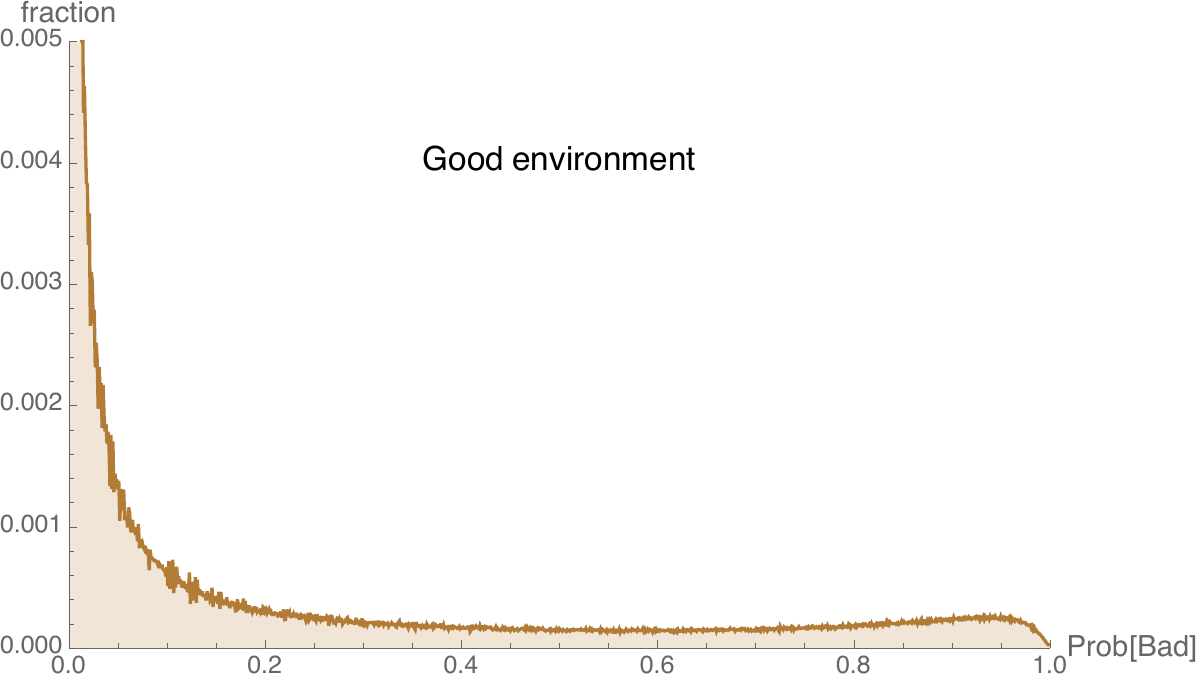}

\caption{Population distribution of subjective probabilities that the environment is bad after 20 time steps, among foxes who have lived that long. When the environment is actually bad (upper panel), all but 4.5 percent of the population accurately come to believe that the environment is more likely to be bad than good. But when the environment is actually good (lower panel), 8.8 percent of the population erroneously come to believe that it is more likely that the environment is bad. All individuals began with a prior probability of 0.5 on the environment being bad.}
  \label{fig:complexmodel}
\end{figure}

\subsection{Population outcomes}

In this signal detection model the fox has two ways to learn about its environment. As before, the fox gains information from exploring a burrow and discovering either a rabbit or badger. But even when the fox chooses not to dig, the fox still gains a small amount of information from observing the cue itself, because the probability of observing a given cue is generally different between the two environments. As a result, individuals will not become stuck forever with an incorrect belief that the environment is bad the way they could in the model of section 2. However, an asymmetry remains between the two kinds of mistakes: it is easier for a fox to learn that it has mistakenly inferred that the environment is good than it is for the fox to learn that it has mistakenly inferred that the environment is bad.

In this model, we observe a qualitatively similar pattern to what we found in the simpler model without cues. Figure \ref{fig:complexmodel} shows the outcome for the whole population when individuals follow the optimal strategy depicted in Figure \ref{fig:complexstrategy}B. When the environment is bad, the majority of foxes correctly learn this. The population distribution of beliefs forms a curve that increases roughly monotonically from left to right, with very few individuals believing that the environment is good, and the great majority correctly believing that the environment is bad. When the environment is good, the majority of foxes learn this as well. But a substantially minority reach the incorrect conclusion that the environment is bad. We see this in the fatter tail of the population distribution of beliefs, and in the existence of a small peak corresponding to the false conclusion that the environment is bad. In this example, roughly twice as many individuals become overly sensitive to loud sounds because they think the environment is bad as become insufficiently sensitive to loud sounds because they think the environment is good (8.8 percent versus 4.5 percent).

One might have thought that having informative cues would always enable the individual to learn to respond appropriately. The reason that it doesn't is that if a fox is in a good environment but is initially unlucky, and receives punishments after observing intermediate cues, then the individual will no longer dig when faced with cues of similar or greater strength. It thus becomes difficult for the fox to correct its mistake and learn that these cues indicate a lower risk of danger than it believes. So this particular fox becomes stuck with an over-sensitivity to the cues of potential danger. Its anxiety level is set too high.
The same thing does not happen when a fox in a bad environment is initially lucky. In that situation, the fox continues to dig at burrows and is soon dealt a harsh punishment by the law of large numbers.

\section{Discussion}

Researchers are discovering many ways in which adaptive behavior can result in seemingly perverse consequences, such as apparent biases or ``irrational" behavior \cite{johnson2013evolution, fawcett2014evolution}. Examples include contrast effects \cite{mcnamara2013adaptive}, state-dependent cognitive biases \cite{harding2004animal, nettle2012evolutionary}, optimism and pessimism \cite{mcnamara2011environmental}, and superstition \cite{foster2009evolution}.

The results of these studies generally explain that the apparently irrational behavior is actually adaptive when understood in its appropriate evolutionary context. In this paper we take a different approach by separating the question of optimal learning rules from the question of whether each individual following such rules ends up behaving optimally. (See Trimmer et al. \cite{trimmer2015depression} for a similar approach applied to clinical depression.)
We show how behavior that is truly dysfunctional (in the sense that it reduces fitness) can arise in a subset of a population whose members follow the optimal behavioral rule, i.e., the rule that generates the highest expected payoff and would thus be favored by natural selection. This approach is well suited to providing insight into
behavioral disorders, since they afflict only a subset of the population and are likely detrimental to fitness. We find that because an exploration-exploitation tradeoff deters further exploration under unfavorable circumstances, optimal learning strategies are vulnerable to erroneously concluding that an environment is bad. A major strength of the model is that it predicts excessive anxiety in a subset of the population, rather than in the entire population as we would expect from ``adaptive defense mechanism'' or  ``environmental mismatch" arguments \cite{nesse2005maladaptation}. 

An interesting aspect of our model is that it predicts the effectiveness of exposure therapy for anxiety disorders \cite{norton2007meta}. In the model, the individuals that are overly anxious become stuck because they no longer observe what happens if they are undeterred by intermediate-valued cues.
If these individuals were forced to take risks in response to the cues that they believe are dangerous but are actually safe, then they would learn that their beliefs were mistaken and would correct their over-sensitivity. This exactly corresponds with the approach employed in exposure therapy. 

Of course such a simple model cannot explain the myriad specific characteristics of real anxiety disorders. One example is that our model fails to capture the self-fulfilling prophecy, or vicious circle aspect, common to excessive anxiety. Being afraid of badgers does not make a fox more likely to encounter badgers in the future. But if a person is nervous because of past failures, that nervousness may be a causal component of future failure. Test anxiety is an example: a student performs poorly on one or more tests, becomes anxious about subsequent tests, and that anxiety contributes to poor performance in the future. Though it is challenging to see how such self-fulfilling anxiety fits into a framework of evolutionary adaptation, modeling the runaway positive feedback aspect of anxiety is an intriguing area for future work.

Another interesting direction for future work would be to investigate the case when the environment varies over time. Our current model is well suited to address a situation in which offspring disperse to different patches in the environment that remain constant over time (that is, when there is spacial variation but not temporal variation). But some environments will also vary over the duration of an individual's lifetime.

Before concluding, we want to point out a consequence in the second model of foxes being able to learn about their environment even when they only observe the cue itself. This means that there are actually two ways that a fox can end up being overly afraid despite living in a good environment. The first parallels our example in the first model: the fox could have had an unlucky early experience with a badger despite detecting only a modest signal, and from this could have mistakenly concluded that it lives in a bad environment. But there is another way that has no analog in the first model: It could be that or fox has never actually encountered a badger firsthand, but rather has received a series of cues more consistent with a bad environment then with a good one, and from these cues alone concluded that he lives in a bad environment even though he's never actually met a badger.

We speculate that these two different scenarios may correspond at least somewhat to different types of anxiety disorders. In the former scenario, present anxiety is the result of past trauma. Post-traumatic stress disorder would appear to be a very straightforward example of such a situation. In the latter, present anxiety would be the result of the mistaken belief that one lives in an unpredictable world, specifically one in which future trauma is difficult to detect and avoid. In both cases, the excessive anxiety on the part of the fox is a consequence of bad luck. But the bad luck can take different forms. In the former case the bad luck comes in the form of a badger observed despite a low signal. In the latter, the bad luck comes in the form of the unsampled signals taking a lower distribution than would be expected given the state of the world.

In general, signal detection models of threat such as these can have a number of moving parts. The degree to which the distribution of cues resulting from good events in bad worlds, and bad events in good worlds happens to overlap is one important factor, and the one we focused on here. Another factor that we've mentioned is when the frequency of good and bad events vary. A further possibility is that the benefits and costs of good and bad events could vary as well. One might even consider mismatch models in which foxes have evolved to distinguish between good and bad worlds but in fact badgers are entirely extinct. Here, the fox might conclude that he lives in a bad world with low discriminability because he has't seen any of the high magnitude signals that he would see in a good world with high discriminability. Considering this range of model possibilities one might be able to demarcate a number of different types of anxiety with different etiology and different predicted forms of treatment. We are currently developing models to explore these possibilities.

In this paper we have illustrated a fundamental design compromise: If an anxiety system is able to learn from experience, even the most carefully optimized system is vulnerable to becoming stuck in a state of self-perpetuating over-sensitivity. This effect is driven by the tradeoff an individual faces between gaining information by experience and avoiding the risk of failure when circumstances are likely unfavorable.
Our results provide a new context for thinking about anxiety disorders: rather than necessarily viewing excessive anxiety as a result of dysregulated or imperfectly adapted neurological systems, we show that many of the features of anxiety disorders can arise from individual differences in experience, even when individuals are perfectly adapted to their environments. We suggest that this phenomenon may be an important causal component of anxiety disorders.

\section*{Acknowledgments}

The authors thank Corina Logan{\bf,} Randy Nesse{\bf, and two anonymous referees} for helpful suggestions and discussions. This work was supported in part by NSF grant EF-1038590 to CTB and by a WRF-Hall Fellowship to FM.

\begin{appendix}
\renewcommand\thefigure{A\arabic{figure}}    
\section*{Appendix A: Sensitivity analysis for Model 1}
\setcounter{figure}{0}    

A central point of this paper is that there is an asymmetry between the fraction of individuals who are wrong about the environment when it is in fact good, and the fraction who are wrong about it when it is bad. In the example we chose in section 2, only 0.2 percent  of the population were optimistic in a bad environment, but 11.1 percent of the population were pessimistic in a good environment. In this appendix we investigate the extent to which changes in the model parameters affect this result.

There are 4 important independent values that parametrize the model. They are: the probability $p_g$ of 
encountering a badger when the environment is good, the probability $p_b$ of encountering a badger when the environment is bad, the discount factor $\lambda$, and the magnitude of the cost of encountering a badger relative to the reward for finding a rabbit, ${C}/{R}$.

\begin{figure}
\includegraphics[width=\linewidth]{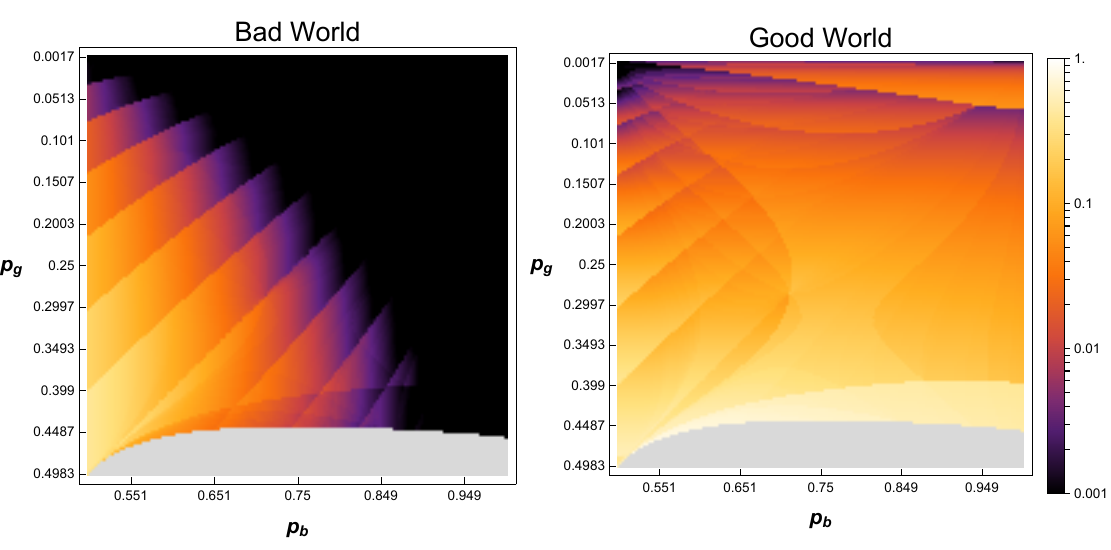}
\begin{center} \includegraphics[width=3.in]{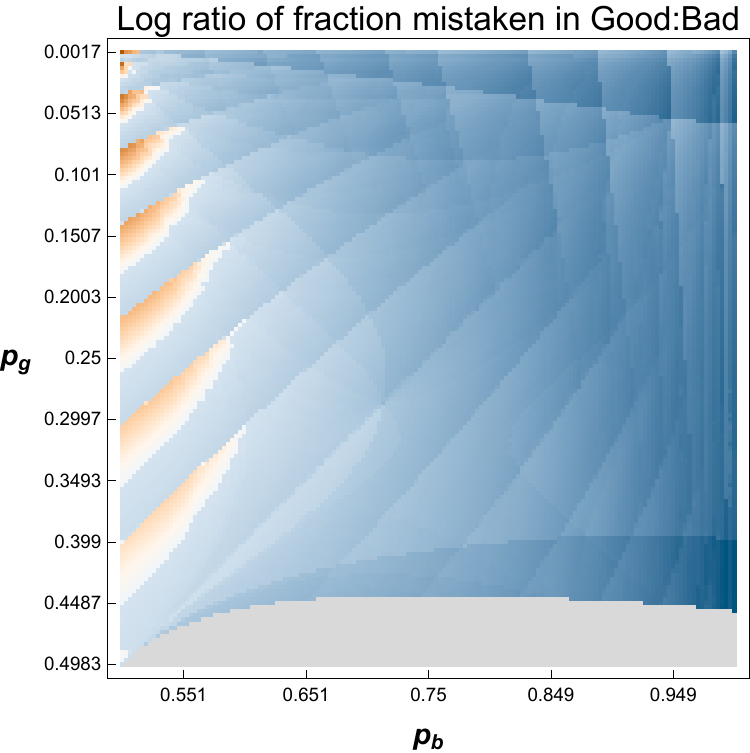} \includegraphics[width=.4in]{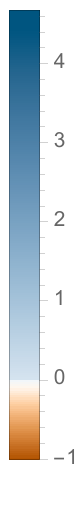}\end{center}

  \caption{Varying the probability of encountering badgers in each environment. With $\lambda=0.95$, $C=1$, and $R=1$, the upper panels show how the fraction of the population that is wrong about the environment varies as a function of the parameters $p_g$ and $p_b$. The upper left shows the fraction that thinks the environment is good when it is actually bad. The upper right panel shows the fraction that thinks the environment is bad when it is actually good. This fraction is measured conditional on survival to the 20th time step, which is the average lifespan when $\lambda = 0.95$. The lower panel illustrates the log (base 10) of the ratio of incorrect inference rates in good and bad environments. For a small set of parameter values (shown in orange), incorrect inferences are more common in the bad environment. The gray area in each plot is a region in which it is not worthwhile to start exploring at all.}
  \label{fig:sen1}
\end{figure}

We first investigate the effect of varying $p_g$ and $p_b$. In order for the state of the environment to matter---for there to be any use of gaining information---we must have the expected payoff be positive when the environment is good, $(1-p_g)R-p_gC>0$, and be negative when the environment is bad, $(1-p_b)R-p_bC<0$. Rearranging these inequalities gives us the constraints
\begin{equation}\label{constraints}
p_b>\frac{R}{R+C}>p_g .
\end{equation}
When $R=C$, as in section 3, these constraints, along with the constraint that $p_g$ and $p_b$ are probabilities that must lie between 0 and 1, restrict us to the square $0.5 < p_b \le 1$, $0 \le p_g < 0.5$. Figure \ref{fig:sen1} displays the results of analyzing the model over a grid of values for $p_g$ and $p_b$ within this square. Plotted is the fraction of the population that is wrong about the environment, as measured after 20 time steps among foxes who have survived that long.

In the upper left panel of Figure \ref{fig:sen1} (bad environment) the fraction of the population that is wrong is negligible everywhere except for 
	the lower left corner of the plot, where the probabilities of encountering a badger in the good environment and in the bad environment are so similar that 20 trials simply does not provide enough information for accurate discrimination. But when the environment is actually good (upper right of Figure \ref{fig:sen1}), it is almost the entire parameter space in which a substantial fraction of the population is wrong about the environment.

Instead of being smooth, the plots are textured by many discontinuities. Optimal behavioral rules cease to explore after small numbers of failures. But these small numbers depend on the parameter values and so discontinuities result around curves in parameter space that are thresholds for different optimal behavioral rules.
However, in spite of the rugged shape of the plot, the basic trend in the upper right-hand panel of Figure \ref{fig:sen1}  is that the fraction of the population that believes the environment is bad when it is actually good increases with $p_g$.
 In the lower panel of Figure \ref{fig:sen1} we see that for over 93 percent of the points in the parameter grid more of the population is wrong in the good environment than in the bad environment. And the small fraction of parameter combinations where this is not the case all occur towards the edge of the parameter space (on the left side in the plot).

\begin{figure}
  \includegraphics[width=\linewidth]{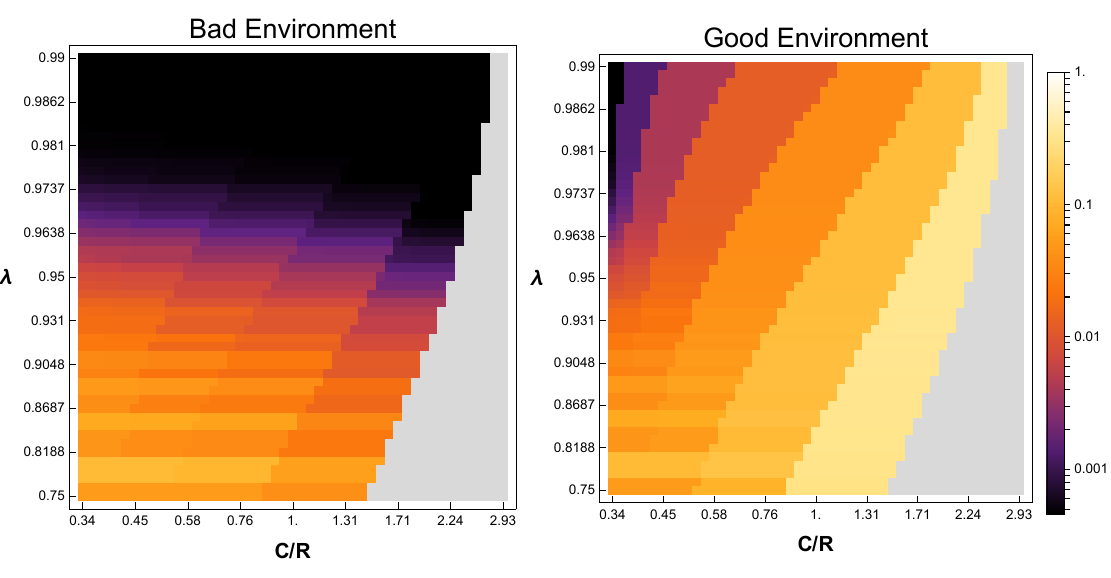}
  \begin{center}
  \includegraphics[width=3in]{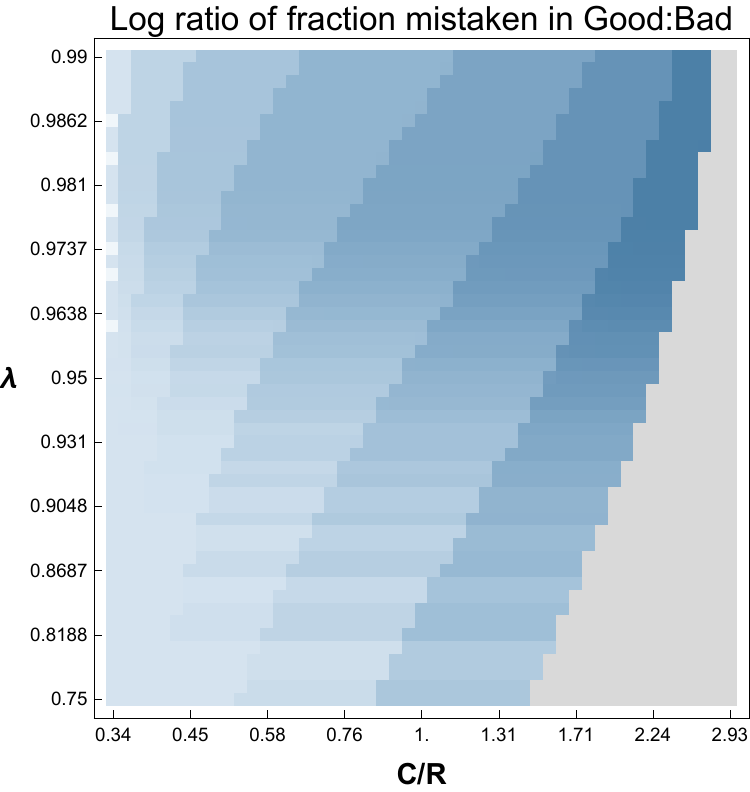}
  \includegraphics[width=.4in]{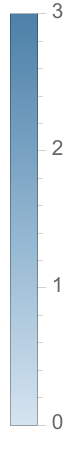}
  \end{center}
  \caption{Varying the discount factor and cost/reward ratio. With $p_g=0.25$ and $p_b=0.75$, the upper panels  show how the fraction of the population that is wrong about the environment varies as a function of $\lambda$ and $C/R$. The upper left plot displays the fraction that thinks the environment is good when it is actually bad; the upper right plot displays the fraction that thinks the environment is bad when it is actually good. This fraction is measured at the time step that is closest to $\frac{1}{1-\lambda}$, the average lifespan given $\lambda$. (The faint horizontal bands towards the lower part of the plots are due to the fact that $\frac{1}{1-\lambda}$ must be rounded to the nearest integer-valued time step.)  The lower plot illustrates the log (base 10) of the ratio of incorrect inference rates in good and bad environments. Here, incorrect inferences are more common in good environments for all parameter values. The gray area in each plot is a region in which it is not worthwhile to start exploring at all. }
  \label{fig:sen2}
\end{figure}

We next investigate the effect of varying the discount factor $\lambda$ and the cost to reward ratio $C/R$, while keeping $p_g$ and $p_b$ constant. Again, for it to matter whether the environment is good or bad, our parameters must satisfy inequalities (\ref{constraints}). Rearranging these gives us the following constraint on the cost/reward ratio: 
\begin{equation}
\frac{1-p_b}{p_b} <\frac{C}{R} < \frac{1-p_g}{p_g}.
\end{equation}
When $p_g = 0.25$ and $p_b = 0.75$ this gives us $\frac{1}{3} < \frac{C}{R} < 3$. Figure \ref{fig:sen2} shows results for the model with values of $\frac{C}{R}$ sampled within this interval and values of $\lambda$ ranging from $0.75$ to $0.99$. The beliefs are measured at the time step that is closest to $\frac{1}{1-\lambda}$, the average lifespan given a discount factor of $\lambda$.

Figure \ref{fig:sen2} shows that, similar to the pattern in Figure \ref{fig:sen1}, the fraction of the population that is wrong when the environment is bad is negligible except when there are not enough time steps in which to make accurate discriminations (in the lower part of Figure \ref{fig:sen2}). By contrast, the fraction of the population that is wrong when the environment is good is non-negligible throughout most of the parameter space. 

Discontinuities due to optimal behavior being characterized by small integer values are especially striking here, especially in the upper right panel of Figure \ref{fig:sen2}. What is happening is that the number of failures it takes before it is optimal to cease to explore is the main outcome distinguishing different parameter choices.  With $p_g$ and $p_b$ fixed, that number also determines the fraction of the population that will hit that number of failures. And so the plot is characterized by a small number of curved bands in which the fraction of the population that is wrong about the environment is nearly constant. Although the value is nearly constant within each band, we can still describe the trend across these bands. We see that the fraction of the population that believes that the environment is bad when it is actually good increases with increasing relative cost or decreasing discount factor.

\section*{Appendix B: Finding optimal behavior}

The signal detection model of section 3, in which the fox uses environmental cues, is defined by the discount factor $\lambda$, the cost of encountering a badger $C$, the reward from catching a rabbit $R$, the initial subjective probability of being in a bad environment $q_0$, the probabilities of badgers in the good environment ($p_g$) and in the bad environment ($p_b$), and the Gaussian distribution parameters
$\mu_{g,c}$, $\sigma_{g,c}$, $\mu_{g,r}$, $\sigma_{g,r}$, $\mu_{b,c}$, $\sigma_{b,c}$, $\mu_{b,r}$, and $\sigma_{b,r}$.
The simpler model of section 2 can be seen as a special case of the more complex model in which the cues carry no information (because the means are all the same). Thus, analyzing the model with cues will also provide an analysis of the simpler model. The problem can be framed as a Markov decision process, and can be analyzed with a dynamic programming approach \cite{bertsekas_dynamic_2005}.


The fox knows the initial prior probability that the environment is bad, and at time step $t$ will also know the outcome of any attempts made before $t$.
For each time step $t$ and all possible previous experience, a behavioral rule specifies the threshold cue level  $u_t$ such that the fox will not dig at the burrow if the observed cue intensity, $x_t$, is greater than $u_t$. The only relevant aspect of previous experience is how this experience changes the current conditional probability $q_t$ that the environment is bad. So an optimally behaving agent will calculate $q_t$ using Bayes' rule, and use this value to determine the threshold level $u_t$. Thus, we can express a behavioral rule as the set of functions $u_t(q_t)$.

Let 
\[
f_{g,c}(x) = \frac{1}{\sigma_{g,c}\sqrt{2\pi}}\,e^{-\frac{(x-\mu_{g,c})^2}{2\sigma_{g,c}^2}}
\]
be the Gaussian distribution function with mean $\mu_{g,c}$ and standard deviation $\sigma_{g,r}$ and similarly let $f_{g,r}(x)$, $f_{b,c}(x)$, and $f_{b,r}(x)$ be the other three corresponding Gaussian distributions.

\subsection*{Expected immediate payoff}

Let $\iota_t$ be the indicator random variable that equals 1 if the burrow contains a badger and equals 0 if the burrow contains a rabbit at time $t$. (Note that the random variables $\iota_t$ and $x_t$ covary.) We now define
$y(q_t, u_t, x_t, \iota_t)$
to be the payoff the fox receives at time $t$ as a function of its threshold ($u_t$), the probability $q_t$ that the environment is bad, and the random variables $x_t$ and $\iota_t$. So

\[
 y(q_t, u_t, x_t, \iota_t) =
  \begin{cases}
   \:\, 0 & \text{if } x_t > u_t \\
   \: R & \text{if } x_t \le u_t \text{ and } \iota_t = 0 \\
   -C & \text{if } x_t \le u_t \text{ and } \iota_t = 1 \\
  \end{cases}
\]
In the bad environment, the probability density of badgers and a cue strength of $x_t$ is $f_{b,c}(x_t)\,p_b$. Likewise, $f_{g,c}(x_t)\,p_g$ gives the probability density of badgers and a cue strength of $x_t$ in the good environment. Similarly $f_{b,r}(x_t)\,(1-p_b)$, and $f_{g,r}(x_t)\,(1-p_g)$ give the same probability densities for rabbits. This allows us to calculate the expected immediate payoff for the strategy of threshold $u_t$ as
\begin{align*}
\mathrm{E}\{y(q_t, u_t, x_t, \iota_t)\} = \int_{-\infty}^{u_t}&\big[\big(f_{b,c}(x_t)\,p_b\,q_t + f_{g,c}(x_t)\,p_g\,(1-q_t)\big)(-C) \\
&+ \big(f_{b,r}(x_t)\,(1-p_b)\,q_t + f_{g,r}(x_t)\,(1-p_g)\,(1-q_t)\big)R\big]\,\mathrm{d}x_t.
\end{align*}

\subsection*{Bayesian updating}

We now describe how the Bayesian probability that the environment is bad, $q_t$, changes with time $t$. That is, we show how $q_{t+1}$ stochastically depends on $q_t$ and the threshold $u_t$.

The probability that the cue intensity is less than or equal to the threshold ($x_t \le u_t$) is given by
\begin{equation*}
\int_{-\infty}^{u_t}\big[\big(f_{b,c}(x_t)\,p_b + f_{b,r}(x_t)\,(1-p_b)\big)\,q_t + \big(f_{g,c}(x_t)\,p_g + f_{g,r}(x_t)\,(1-p_g)\big)(1-q_t)\big]\,\mathrm{d}x_t.
\end{equation*}
The probability that the cue intensity is greater than the threshold ($x_t > u_t$) is the complement (the above quantity subtracted from 1).

If $x_t > u_t$, then the fox does not explore the burrow, but gains information about the environment from the cue itself, $x_t$. In the bad environment, the probability density on cues $x$ is given by
\[
p_b\,f_{b,c}(x) + (1-p_b)f_{b,r}(x).
\]
Similarly, in the good environment it is given by
\[
p_g\,f_{g,c}(x) + (1-p_g)f_{g,r}(x).
\]
So according to Bayes' rule, the posterior distribution on $q_{t+1}$ given an $x_t > u_t$ is
\[
q_{t+1} = \frac{q_t\,\big(p_b\,f_{b,c}(x_t) + (1-p_b)f_{b,r}(x_t)\big)}{q_t\,\big(p_b\,f_{b,c}(x_t) + (1-p_b)f_{b,r}(x_t)\big) + (1-q_t)\,\big(p_g\,f_{g,c}(x_t) + (1-p_g)f_{g,r}(x_t)\big)}.
\]

On the other hand if $x_t \le u_t$, the fox decides to dig. The fox will then observe both the cue and whether the burrow contains a badger or a rabbit. The conditional probability that the burrow contains a badger given the cue $x_t$ and a bad environment is 
\begin{align*}
\mathrm{P}(\text{badger}\,|\,x_t, \text{bad env.}) &= \frac{\mathrm{P}(x_t\,|\,\text{badger},\text{bad env.})\,\mathrm{P}(\text{badger}\,|\,\text{bad env.})}{\mathrm{P}(x_t\,|\,\text{bad env.})}\\
&= \frac{f_{b,c}(x_t)\,p_b}{f_{b,c}(x_t)\,p_b + f_{b,r}(x_t)\,(1-p_b)}.
\end{align*}
(Note that technically some of these quantities are probability densities rather than probabilities.) Similarly, if the environment is good, then
\begin{align*}
\mathrm{P}(\text{badger}\,|\,x_t, \text{good env.}) &= \frac{\mathrm{P}(x_t\,|\,\text{badger},\text{good env.})\,\mathrm{P}(\text{badger}\,|\,\text{good env.})}{\mathrm{P}(x_t\,|\,\text{good env.})}\\
&=\frac{f_{g,c}(x_t)\,p_g}{f_{g,c}(x_t)\,p_g + f_{g,r}(x_t)\,(1-p_g)}.
\end{align*}
Thus, the total probability of encountering a badger ($\iota_t=1$) given an $x_t \le u_t$ is
\[
q_t\,\frac{f_{b,c}(x_t)\,p_b}{f_{b,c}(x_t)\,p_b + f_{b,r}(x_t)\,(1-p_b)} + (1-q_t)\frac{f_{g,c}(x_t)\,p_g}{f_{g,c}(x_t)\,p_g + f_{g,r}(x_t)\,(1-p_g)}.
\]

If the burrow does contain a badger, then by Bayes' rule we can express the posterior probability that the environment is bad as follows.
\begin{align*}
\mathrm{P}(\text{bad env.}\,|\,x_t, \text{badger}) &= \frac{\mathrm{P}(x_t, \text{badger}\,|\,\text{bad env.})\,\mathrm{P}(\text{bad env.})}{\mathrm{P}(x_t, \text{badger})} \\
\mathrm{P}(\text{bad env.}\,|\,x_t, \text{badger}) &= \frac{\mathrm{P}(x_t\,|\,\text{badger},\text{bad env.})\,\mathrm{P}(\text{badger}\,|\,\text{bad env.})\,\mathrm{P}(\text{bad env.})}{\mathrm{P}(x_t, \text{badger})} \\
q_{t+1} &= \frac{f_{b,c}(x_t)\,p_b\,q_t}{f_{b,c}(x_t)\,p_b\,q_t + f_{g,c}(x_t)\,p_g\,(1-q_t)}.
\end{align*}
Perfectly analogous calculations hold for the case when the burrow contains a rabbit ($\iota_t = 0$).

Below, we will express $q_{t+1}$ as a function,
\begin{equation*}
q_{t+1} = w(q_t, u_t, x_t, \iota_t),
\end{equation*}
that depends on the threshold $u_t$, the probability $q_t$, and the random variables $x_t$ and $\iota_t$, as described above.

\subsection*{Dynamic programming}

The dynamic programming algorithm now consists of recursively calculating the maximum payoff attainable over all time steps subsequent to $t$, as a function of the current probability that the environment is bad. This maximum payoff is denoted $V_t(q_t)$, and the recursive formula is
\[
V_t(q_t) = \max_{u_t} \mathrm{E}\{y(q_t, u_t, x_t, \iota_t) + \lambda\,V_{t+1}(w(q_t, u_t, x_t, \iota_t))\}.
\]
And the optimal decision rule functions, $u_t^*$, are given by
\[
u_t^*(q_t) = \operatorname*{arg\,max}_{u_t} \mathrm{E}\{y(q_t, u_t, x_t, \iota_t) + \lambda\,V_{t+1}(w(q_t, u_t, x_t, \iota_t))\}.
\]

Because $q_t$ is a continuous variable, a discrete approximation must be used for the actual computation. Then the table of values for $V_{t+1}$ is used to compute the values for $V_t$, indexed by $q_t$. For $q_t$ we used 1001 discrete values ($0, 0.001, 0.002, \ldots, 1$).
As mentioned in section 3, we set $\mu_{g,c} = 2, \mu_{b,c} = 1$, $\mu_{g,r} = \mu_{b,r} = 0$, and $\sigma_{g,r} = \sigma_{g,c} = \sigma_{b,r} = \sigma_{b,c} = 0.5$. To discretize $x_t$, we must pick minimum and maximum values, which we set at $-2$ and $4$, respectively. Within this interval we discretized $x_t$ to 200 values. Because there is a tiny area lost at the ends of the distributions, we renormalized the total probabilities to 1.

The algorithm then gives us two tables: one containing the expected values and the other containing the optimal decision rule functions, or thresholds, $u_t^*(q_t)$, which are indexed by our grid of values for $q_t$. To find the optimal behavior in the limit as the possible lifetime extends towards infinity, the recursion is repeated until the optimal decision rules converge \cite{bertsekas_dynamic_2005}. The algorithm was implemented in python.

Once we have found the optimal decision rule, for each time step we can calculate the expected proportion of the population that has each value of $q_t$ as its estimate. Since the behavioral rule specifies the threshold for each value of $q_t$, we can use the distribution derived above for $q_{t+1} = w(q_t, u_t, x_t, \iota_t)$ to calculate the proportions for time $t+1$ given the proportions for time $t$. Because we discretized the $q_t$ values, we round the calculation of $q_{t+1}$ to the nearest one thousandth.

\end{appendix}
\newpage
\bibliographystyle{unsrt}
 \bibliography{Anxiety}
\end{document}